\documentclass{emulateapj}
\shorttitle{SN~2005ap}
\shortauthors{Quimby et al.}
\usepackage{epsfig}
\usepackage{apjfonts}
\begin{document}

\title{SN~2005ap: A Most Brilliant Explosion}

\author{
  Robert M. Quimby\altaffilmark{1},
  Greg Aldering\altaffilmark{2},
  J. Craig Wheeler\altaffilmark{1},
  Peter H\"{o}flich\altaffilmark{3},
  Carl W. Akerlof\altaffilmark{4},
  Eli S. Rykoff\altaffilmark{4}
}

\altaffiltext{1}{
  Department of Astronomy,
  University of Texas,
  Austin, TX 78712, USA
}
\altaffiltext{2}{
  Physics Division,
  LBNL,
  1 Cyclotron Road,
  Berkeley, CA 94720, USA
}
\altaffiltext{3}{
  Department of Physics,
  Florida State University,
  Tallahassee, FL 32312, USA
}
\altaffiltext{4}{
  University of Michigan,
  2477 Randall Laboratory, 450
  Church Street, Ann Arbor, MI, 48104, USA
}

\begin{abstract}

We present unfiltered photometric observations with ROTSE-III and
optical spectroscopic follow-up with the HET and Keck of the most
luminous supernova yet identified, SN~2005ap. The spectra taken about
3 days before and 6 days after maximum light show narrow emission
lines (likely originating in the dwarf host) and absorption lines at a
redshift of $z=0.2832$, which puts the peak unfiltered magnitude at
$-22.7 \pm 0.1$ absolute. Broad P-Cygni features corresponding to
H$\alpha$, \ion{C}{3}, \ion{N}{3}, and \ion{O}{3}, are further
detected with a photospheric velocity of $\sim 20,000$\,km\,s$^{-1}$.
Unlike other highly luminous supernovae such as 2006gy and 2006tf that
show slow photometric evolution, the light curve of SN~2005ap
indicates a 1-3 week rise to peak followed by a relatively rapid
decay. The spectra also lack the distinct emission peaks from
moderately broadened (FWHM $\sim 2,000$\,km\,s$^{-1}$) Balmer lines
seen in SN~2006gy and SN~2006tf. We briefly discuss the origin of the
extraordinary luminosity from a strong interaction as may be expected
from a pair instability eruption or a GRB-like engine encased in a
H/He envelope.

\end{abstract}


\section{Introduction}

Luminous supernovae (SNe) are most commonly associated with the Type
Ia class, which are thought to involve explosions of white dwarf
stars; however, the brightest SNe (those with absolute magnitudes $M <
-20$) are all associated with the deaths of stars that begin their
lives with main sequence masses $M_{\rm{ms}} > 7-8\,M_{\odot}$. The
current record holder, SN~2006gy, is thought to be an explosion of a
supermassive star ($M_{\rm{ms}} \sim 100-150\,M_{\odot}$) and it may
represent the first detection of a supernova triggered by pair
instability~\citep{smith2007}. The remaining bright SNe
are attributed to core-collapse induced explosions (CCSNe).

CCSNe are classified as Type~Ib/c if they lack strong hydrogen lines
in their spectra and Type~II otherwise~\citep{filippenko1997}. The
latter class has been partitioned into three groups: (1) Type~IIn show
narrow emission lines in their spectra, (2) Type~II-L have linearly
declining light curves, and (3) Type~II-P are ``normal'' hydrogen rich
events exhibiting a slow photometric evolution phase (i.e. a
plateau). SN~1993J is sometimes classified as a ``Type~IIb'' as its
spectral characteristics evolved from Type~II-like to resemble a
Type~Ib, which lack H but show He lines (unlike the Type Ic
class). CCSNe may span a continuum with the different spectral types
explained in terms of decreasing envelope masses (II-IIb-Ib-Ic).

Long Duration gamma-ray bursts (GRBs) are occasionally observed to
exhibit SN features. These appear as bumps in the optical afterglow
light curves~\citep{bloom99,lazzati01}, or more revealingly as broad
spectral features resembling SNe~Ic that emerge in the spectra of
afterglows 1-2 weeks after the burst~\citep{stanek03}. Given the
connection of GRBs to SNe~Ic and the continuum of CCSNe spectral
types, it is natural to consider the observational signature of a GRB
engine erupting within an envelope of some mass
(e.g. \citealt{macfadyen2001}).
Such material could slow the ultra-relativistic flow and thus mask the
gamma-ray beacon announcing their creation, unlike their stripped
progenitor cousins. \citet{young2005} have explored this possibility
and constructed models that can explain the light curves of the bright
subclass within the Type~II-L group.

In this letter we report the discovery of SN~2005ap, a transient
optical source that peaked at about $-22.7$ magnitude. We present the
photometry in \S\ref{phot}, the spectral observations and modeling in
\S\ref{spec}, and we offer discussion and conclusions in
\S\ref{conclusions}. SN~2005ap is the brightest supernova ever
identified and may shed light on energy production mechanisms in
cosmic explosions.

\section{Photometry}\label{phot}
\begin{figure}
  \centerline{\epsfig{file=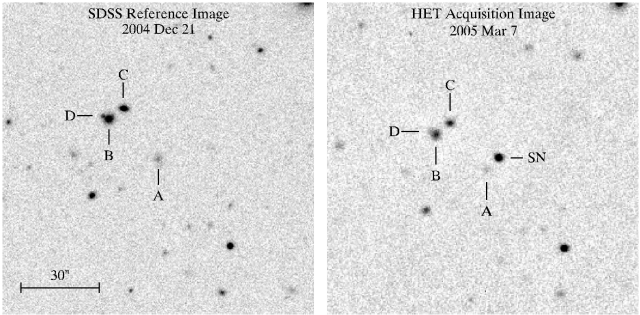,width=\columnwidth}}
  \caption{SDSS reference image (left) and HET/LRS acquisition image
    of SN~2005ap (right). The SDSS image is a combination of the $g$-,
    $r$-, and $i$-band frames taken 80 days before maximum light. The
    host is not clearly detected on the combined SDSS frame, but four
    neighboring galaxies are labeled A-D. The SN is marked on the
    HET/LRS image.\label{fig:chart}}
\end{figure}

\begin{figure}
  \centerline{\epsfig{file=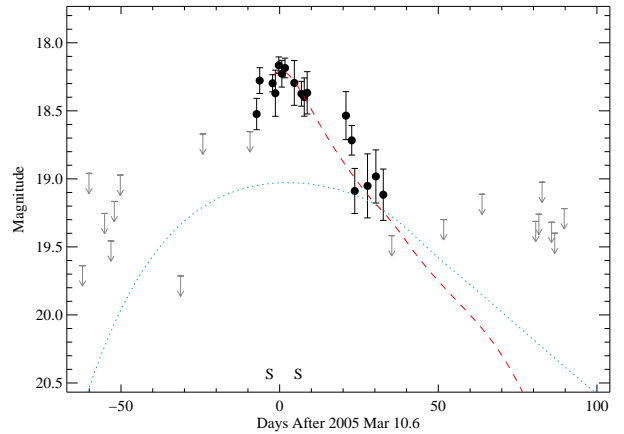,width=\columnwidth}}
  \caption{ROTSE-III unfiltered light curve of SN~2005ap calibrated
    against the USNO-B1.0 R2 magnitudes. Detections are indicated by
    filled circles.  Upper limits fainter than 18.5 mag prior to
    discovery and 19.0 after the last detection are marked with
    arrows.  The red dashed line depicts the $V$-band light curve of a
    SN~II-L (SN~1980K [\citealt{barbon1982,buta1982}]) dilated in time
    by $1+z$ and arbitrarily shifted by 4 magnitudes to match the
    peak. The dotted blue line is the unfiltered light curve of
    SN~2006gy from \citet{smith2007}, corrected for its 1.68
    magnitudes of extinction and phased to match the peak of
    SN~2005ap, as it would appear at a redshift of $z=0.2832$. The HET
    and Keck spectral epochs are each marked along the abscissa with
    an ``S.''\label{fig:lc}}
\end{figure}

SN~2005ap was discovered on unfiltered optical images taken with the
$0.45\,m$ ROTSE-IIIb (Robotic Optical Transient Search Experiment)
telescope~\citep{akerlof03}, which is located at the McDonald
observatory in west Texas. The transient was found on 2005 March 3
(UT dates are used throughout this letter) in images taken in the
course of the Texas Supernova Search (TSS;
\citealt{quimby_phd}). SN~2005ap was identified in a field centered on
the Coma Galaxy Cluster after removal of static sources via a modified
version of the image subtraction code developed by the Supernova
Cosmology Project~\citep{perlmutter1999}. SN~2005ap is located at
$\alpha=13^h01^m14\fs83$, $\delta=+27\arcdeg43\arcmin32\farcs3$
(J2000), which is $6.2''$ projected from SDSS~J130115.12+274327.5
(galaxy A in Fig.~\ref{fig:chart}); however, \citet{adami2006} have
cataloged a source $0.6'' \pm 0.3''$ west of the transient with
$B=24.6 \pm 0.2$, $V=23.9 \pm 0.2$, and $R=23.5 \pm 0.3$, and we
identify this as the host of SN~2005ap.  SN~2005ap reached an
unfiltered magnitude of $18.13 \pm 0.08$ (calibrated against the
USNO-B1.0 R2
values\footnote{\url{http://www.nofs.navy.mil/data/fchpix/}}) on March
10.28 and was observed at a similar brightness the following two
nights (see Fig.~\ref{fig:lc}). The transient was detected over a 40
day period, and it has since remained undetected over two years of
observations by the ROTSE-III telescopes (170 nights with limiting
magnitudes $18 < M_{\rm lim} < 19.5$). A 4$^{\rm{th}}$ order
polynomial fit to the detections shows maximum light occurred on 2005
March 10.6. SN~2005ap brightened over at least a 7 day period to reach
its peak. The shape of the light curve is consistent with supernova
templates that take $\sim 3$ weeks to reach maximum light.

\section{Spectroscopy and Spectral Fitting}\label{spec}

\begin{figure}
  \centerline{\epsfig{file=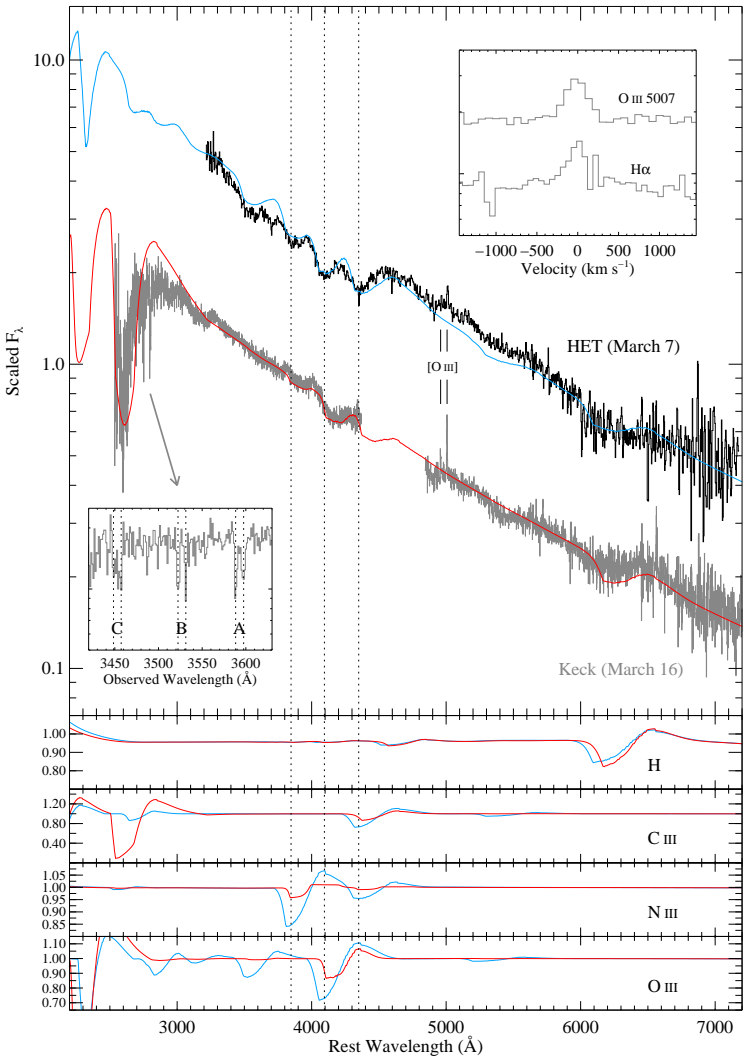,width=4in}}
  \caption{HET/LRS spectrum of SN~2005ap taken 3 days before maximum
    light (black curve) and the Keck/LRIS data from 6 days after
    maximum light (grey curve). The blue and red curves show spectral
    fits to the data from the highly parameterized SYNOW code. An
    unidentified P-Cygni profile is seen around 3200\,\AA. The left
    inset is a detail from the Keck spectra showing three narrow
    absorption doublets that we identify as intervening
    \ion{Mg}{2}\,$\lambda$2796,2804 systems. The upper inset shows the
    narrow H$\alpha$ and [\ion{O}{3}]\,$\lambda$5007 from the Keck
    data plotted in velocity space relative to the $z=0.2832$
    restframe. The bottom panels show the individual contributions
    from the four ions included in the fits.\label{fig:spec}}
\end{figure}

A spectrum of SN~2005ap acquired 2005 March 7 with the Low Resolution
Spectrograph (LRS; \citealt{hill1998}) on the Hobby-Eberly Telescope
(HET) shows a very blue continuum with subtle line features including
broad absorption and at least one narrow emission line
(Fig.~\ref{fig:spec}). Keck/LRIS~\citep{oke1995} data obtained 9 days
later exhibit similar behavior, and additionally show narrow
absorption lines at wavelengths not covered by the HET. We identify
the narrow emission lines observed around 6363 and 6425\,\AA\ as
[\ion{O}{3}]\,$\lambda$4959,5007, and a narrow emission line around
8422\,\AA\ in the Keck spectrum as H$\alpha$. The narrow absorption
doublet seen at 3595\,\AA\ corresponds to
\ion{Mg}{2}\,$\lambda2796,2803$ in this same frame, and sets a firm
lower limit for the redshift of SN~2005ap at $z=0.2832$. At this
redshift, the peak unfiltered magnitude is $-22.7 \pm 0.1$ absolute
($\Omega_{m}=0.265$, $\Omega_{\Lambda}=0.735$,
$H_0=71$\,km\,s$^{-1}$\,Mpc$^{-1}$). Corrections for minor
($A_R=0.022^{m}$) Galactic extinction~\citep{schlegel1998} and
absorption within the host are not included.

The alignment of the slit for the HET data includes the southern edge
of galaxy B, which is $25.7''$ from the SN
(Fig.~\ref{fig:chart}). These spectra indicate that the redshift of
galaxy B is consistent with the \ion{Mg}{2} absorption doublet at
3525\,\AA\ ($z \sim 0.26$; projected separation $\sim 100$\,kpc). The
remaining \ion{Mg}{2} absorption system is consistent with the SDSS
photometric redshift of galaxy C in Figure 1, to within the
uncertainty~\citep{sdssdr5}.

\begin{figure}
  \centerline{\epsfig{file=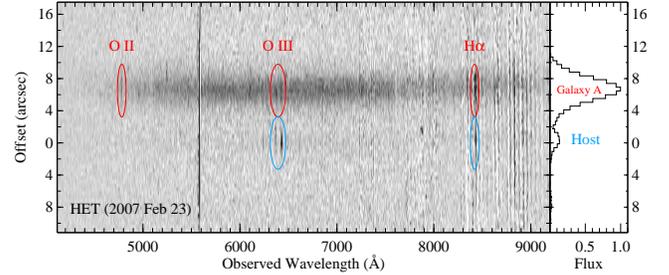,width=\columnwidth}}
  \caption{HET/LRS spectra taken about two years after SN~2005ap faded
    below the detection limit of ROTSE-III. The slit was oriented to
    include the location of the SN as well as galaxy A. The narrow
    [\ion{O}{3}]\,$\lambda$4959,5007 doublet redshifted by $z=0.283$
    is still clearly present (larger blue circle), suggesting this
    emission was not associated with SN~2005ap. Galaxy A also exhibits
    the [\ion{O}{3}] doublet as well as [\ion{O}{3}]\,$\lambda$3727
    (red circles). H$\alpha$ is detected in both galaxies, although it
    is confused with the imperfectly subtracted night sky lines. In
    addition to the narrow emission lines, the host is detected as a
    weak continuum source. The integrated light is plotted in the
    right panel. Note the host emission is distinct from galaxy
    A.\label{fig:spec2d}}
\end{figure}

A second observation was conducted with the HET on 2007 February 23,
long after the transient had faded below the detection limit of
ROTSE-III. The slit was aligned to include the position of SN~2005ap
as well as galaxy A. The [\ion{O}{3}]\,$\lambda$4959,5007 doublet is
clearly detected near the location of the transient
(Fig.~\ref{fig:spec2d}). This may indicate that the narrow emission
lines are not directly associated with SN~2005ap, but rather are
intrinsic to the faint host, which has an absolute magnitude, $M_R$,
of about -16.8. Like the host, galaxy A is also redshifted by $z \sim
0.283$, giving it an absolute magnitude of about $M_R=-20.4$ and the
pair a projected separation of 28.3\,Kpc.

We used the parameterized spectral synthesis code,
SYNOW~\citep{jeffery_branch1990}, to identify broad features in the
spectra of SN~2005ap. First we note that if the continuum is a black
body, then the steep dependence with wavelength implies a 16,000 to
20,000\,K source. Alternatively, the continuum may be a power-law,
$f_{\nu} \propto \nu^{\beta}$, with $\beta \sim 0.7$. We paid special
attention to the extraction of the blue channel Keck data. The
rollover at the blue end cannot signal a black body peak within the
allowed temperature range. The spectra are modeled as 20,000\,K and
17,000\,K black body continua for the March 7 and 16 data,
respectively, with superposed P-Cygni profiles.

The lack of distinct features in the 4500 to 8000\,\AA\ range imposes
significant constraints on the cause of the broad absorption features
between rest 3500 and 4500\,\AA.  We obtain acceptable fits to both
spectral epochs from SYNOW using only \ion{H}{1}, \ion{C}{3},
\ion{N}{3}, and \ion{O}{3} (colored lines in
Fig.~\ref{fig:spec}). \ion{He}{2} may also be present, but its
strongest feature is blended with \ion{C}{3}/\ion{N}{3} and is not
strictly required. The presence of such high excitation species is
consistent with the high temperatures implied by the blue continuum
slope. SYNOW indicates the \ion{C}{3} feature at about 2600\,\AA\ is
temperature dependent, and \ion{C}{3} can acceptably fit the rollover
at the blue end of the Keck spectrum if the ion temperature is lowered
to about 7,500 K (the apparent inconsistency between the ion and
continuum temperatures may arise due to assumptions in the SYNOW code,
which neglects non-LTE effects).  Between March 7 and March 16 the
lines shift redward by 50-60\,\AA\ (note the HET and Keck wavelength
solutions agree to $\sim 0.1\%$). This indicates a significant
deceleration of the photosphere between the two epochs. The SYNOW fits
give a photospheric velocity of $\sim$23,000\,km\,s$^{-1}$ on March 7,
decreasing to $\sim$19,000\,km\,s$^{-1}$ on March 16.

\section{Discussion and Conclusions}\label{conclusions}

We have presented photometric and spectroscopic observations of the
Type~II supernova 2005ap. Narrow absorption and emission lines from
within the dwarf host indicate a redshift of $z=0.2832$. The
successful SYNOW fits and the lack of additional narrow absorption
systems indicate the SN cannot be far beyond this redshift. Correcting
only for distance, the peak unfiltered magnitude was $-22.7 \pm 0.1$
absolute, which corresponds to $\sim 4 \times 10^{44}$\,erg\,s$^{-1}$
assuming no bolometric correction. We consider below the possibility
that SN~2005ap was an AGN or a GRB afterglow, but we find it is most
consistent with a supernova explosion. We conclude SN~2005ap is the
most luminous supernova ever identified, and roughly twice as bright
as the previous record holder, SN~2006gy.

Although the peak absolute magnitude of SN~2005ap is comfortably in
the range $-21 > M > -23$ typical of X-ray selected AGN around this
redshift~\citep{anderson2007}, this interpretation is unlikely. First
and foremost, this transient occurred in a region of the sky that has
been exceptionally well observed across the electromagnetic spectrum
for decades; no indication of a possible AGN at this position has
previously been uncovered. Specifically, there are no X-ray detections
identified in either the ROSAT All-Sky Survey source
catalogs~\citep{voges1999,voges2000} or the XMM-Newton Serendipitous
Source
Catalogue\footnote{\url{http://xcatdb.u-strasbg.fr/xcatdb-corr/}}, nor
is there a cataloged radio source present in either the NRAO VLA Sky
Survey~\citep{condon1998} or the FIRST survey
catalog~\citep{white1997}. Second, the spectra are quite different
from normal AGN, showing no H$\beta$ or \ion{Mg}{2} emission features.
There is some similarity to the featureless spectra of certain
blazars, although blazars lack P-Cygni line profiles and often show
narrow lines from giant elliptical hosts~\citep{wurtz1996}. Third, the
best position for SN~2005ap is offset from its host at the 2\,$\sigma$
level, and at peak the transient was more than 5 magnitudes brighter
than the underlying host light. Finally, no repeat outbursts have been
detected at this location in two years of monitoring.

The host galaxy is reminiscent of settings for long duration
GRBs~\citep{fruchter2006}; however, the observed population of GRB
afterglows typically decline as power-laws ($f \propto t^{\alpha}$
with $\alpha \sim -1$), and the 1-3 week rise and slow fading of
SN~2005ap are inconsistent with this trend. An off-axis orphan
afterglow is a interesting possibility, although current models
predict a rapid power-law decline after maximum light with $\alpha
\sim -1.5$~\citep{nakar_piran2003}, which does not appear to match the
observations.

Except for its unusually bright peak magnitude, the observations of
SN~2005ap are quite consistent with the typical behavior of
supernovae. The time scales for the photometric evolution are common
among SNe, as is the presence of broad, P-Cygni profiles in the
spectra. There have been SNe with continua slopes about as blue as
SN~2005ap (e.g. SN~1998S; \citealt{fassia2001}), but the existence of
\ion{C}{3}, \ion{N}{3}, and \ion{O}{3} features is unprecedented. The
broad H$\alpha$ P-Cygni profile is of course the defining feature of
SNe~II. With no narrow emission lines clearly associated with the
explosion and lacking an observed photometric plateau phase, SN~2005ap
could be classified as a Type~II-L. The light curve is similar in
shape to the behavior of SNe~II-L (Fig. \ref{fig:lc}) although it is
3-4 mag too bright. The spectra are also roughly similar to the early
observations of SN~1979C~\citep{branch1981}, except SN~2005ap shows a
much bluer continuum. However, SNe~II-L show broad, asymmetric
H$\alpha$ emission with little to no blueshifted absorption, and they
also have H$\beta$, H$\gamma$, and \ion{He}{1}/\ion{Na}{1} P-Cygni
profiles. These characteristics are wanting in the observations of
SN~2005ap. While most SNe~II-L have narrowly distributed peak
magnitudes, a small subsample (including SN~1979C) deviate to brighter
values~\citep{gaskell1992}. \citet{young2005} have suggested these
deviants are powered by a GRB engine within a H/He envelope.
 
The luminosity of normal SNe~II is powered by energy deposited by the
explosion shock, but this requires that the initial radius of the
progenitor be significant compared to the radius of the photosphere in
order to minimize adiabatic losses. For an exceptionally bright object
like SN~2005ap, this would require an unrealistically large
radius. With a photosphere moving at about v$_{ph} \sim
20,000$\,km\,s$^{-1}$ and a time of maximum light of t$_{max} \sim
1\times10^6$\,s, the radius of the photosphere near maximum light is
R$_{ph}\sim 2\times10^{15}$\,cm, which is much too large to correspond
to a standard red supergiant.  

If the luminosity arises in the collision of the ejecta with a
surrounding, perhaps dense, shell of circumstellar matter shed by a
wind or a process like an LBV mass ejection, then we might see the
radiation emitted by a shocked, thermalized, shell. For a shell of
mass M$_{sh}$, radius R, thickness $\Delta$R, and optical depth,
$\tau$, maximum light will occur when the diffusion time, t$_{diff}
\sim 3\Delta R \tau$/c, is comparable to the dynamical time, t$_{dyn}
\sim \Delta$R/v$_{ph}$. This gives the optical depth at maximum light,
$\tau_{max} \sim 1/3$(c/v$_{ph}$)$\sim 5$ neglecting any differences
between the photospheric velocity and the mean ejecta velocity. Since
M$_{sh} \sim 4 \pi$ R$^{2} \tau_{max}/\kappa$, the mass of the shell
would be about 1.3\,M$_{\odot}$, with $\kappa = 0.2$ for electron
scattering. The peak luminosity would be
L$\sim$1/2\,M$_{sh}$\,v$_{ph}^2$/t$_{max}$ $\sim
2\times10^{45}$\,erg\,s$^{-1}$, more than directly observed, but
roughly consistent with the observations.
The model for SNe~II-L by \citet{young2005} is a variation on this
theme, but with the optical luminosity produced by non-thermal
emission from the afterglow process.

Another possibility is that a separate internal source of energy
powers the luminosity.  A model driven purely by radioactive decay
would require a large asymmetry to prevent the nickel mass (10's of
M$_{\odot}$) from exceeding the ejecta mass~\citep{hww1999} , which
must be modest to account for the rapid rise and decline of the light
curve. Whether this could be accomplished with a large nickel mass
suggesting a pair formation event, or with a lower nickel mass more
commensurate with a ``normal" core collapse explosion requires more
extended analysis.
                                                                                
\citet{maeda2007} propose the rapid spin-down of a magnetar to produce
a high intrinsic luminosity and subsequent decline from the second
peak of SN~2005bf. It is possible that an appropriate choice of ({\it
ad hoc}) parameters could produce the peak we observe in SN~2005ap,
but with a smaller mass and larger temporal decay index to account for
the brighter, faster peak.

In two years of operation, the Texas Supernova Search has uncovered
the most luminous supernova (this work), the second most luminous
(SN~2006gy [\citealt{smith2007,ofek2007}]), and a third, highly
luminous event (SN~2006tf [\citealt{SN2006tf_iauc}]). The volume
monitored over this time (roughly matching the total from all past
narrow-field CCD surveys targeting known galaxies), the high
completeness of the sample resulting from a tight (nightly) cadence
and ability to work in the cores of galaxies, along with prompt high
S/N spectroscopic follow-up, have been key to finding and fully
classifying these events. The lack of SN~2005ap-like events reported
in previous surveys may suggest they are intrinsically rare, but the
lack of SN~2006gy-like events might be understood if they are
preferentially located in the cores of bright galaxies. Our discovery
of new classes of bright transients bodes well for more extensive
future surveys of the time-variable sky.

\acknowledgments We would like to thank the staff of the Hobby-Eberly
Telescope and McDonald Observatory for their support. We give specific
thanks to V. Riley and B. Roman for their observations with the HET,
and to F. Castro, P. Mondol, and M. Sellers for their efforts in
screening potential SN candidates. We thank M. Kowalski and D.
Schlegel for their assistance with the Keck observations, and
D. Branch and J. Parrent for discussions on the SYNOW code. This
research is supported, in part, by NASA grant NAG 5-7937 (PH), NSF
grants AST0307312 (PH) and AST0406740 (RQ \& JCW), and DOE contract
DE-AC02-05CH11231 (GA). The authors wish to recognize and acknowledge
the very significant cultural role and reverence that the summit of
Mauna Kea has always had within the indigenous Hawaiian community.  We
are most fortunate to have the opportunity to conduct observations
from this mountain.


\begin{thebibliography}{35}
\expandafter\ifx\csname natexlab\endcsname\relax\def\natexlab#1{#1}\fi

\bibitem[{{Adami} {et~al.}(2006){Adami}, {Picat}, {Savine}, {Mazure}, {West},
  {Cuillandre}, {Pell{\'o}}, {Biviano}, {Conselice}, {Durret}, {Gallagher},
  {Gregg}, {Moreau}, \& {Ulmer}}]{adami2006}
{Adami}, C., {Picat}, J.~P., {Savine}, C., {et~al.} 2006, \aap, 451, 1159

\bibitem[{{Adelman-McCarthy} {et~al.}(2007){Adelman-McCarthy}, {Agueros},
  {Allam}, {Anderson}, {Anderson}, {Annis}, {Bahcall}, {Bailer-Jones},
  {Baldry}, {Barentine}, {Beers}, {Belokurov}, {Berlind}, {Bernardi},
  {Blanton}, {Bochanski}, {Boroski}, {Bramich}, {Brewington}, {Brinchmann},
  {Brinkmann}, {Brunner}, {ari}, {Carey}, {Carliles}, {Carr}, {Castander},
  {Connolly}, {Cool}, {Cunha}, {Csabai}, {Dalcanton}, {Doi}, {Eisenstein},
  {Evans}, {Evans}, {Fan}, {Finkbeiner}, {Friedman}, {Frieman}, {Fukugita},
  {Gillespie}, {Gilmore}, {Glazebrook}, {Gray}, {Grebel}, {Gunn}, {Haas},
  {Hall}, {Harvanek}, {Hawley}, {Hayes}, {Heckman}, {Hendry}, {Hennessy},
  {Hindsley}, {Hirata}, {Hogan}, {Hogg}, {Holtzman}, {Ichikawa}, {Ichikawa},
  {Ivezic}, {Jester}, {Johnston}, {Jorgensen}, {c}, {mann}, {Kent}, {Kleinman},
  {Knapp}, {Kniazev}, {Kron1}, {Krzesinski}, {Kuropatkin}, {Lamb}, {Lampeit},
  {Lee}, {Leger}, {Lima}, {Lin}, {Long}, {Loveday}, {Lupton}, {Mandelbaum},
  {Margon}, {Martinez-Delgado}, {Matsubara}, {McGehee}, {McKay}, {Meiksin},
  {Munn}, {Nakajima}, {Nash}, {Jr.}, {Newberg}, {Nichol}, {Nieto-Santisteban},
  {Nitta}, {Oyaizu,}, {Okamura}, {Ostriker}, {Padmanabhan}, {Park}, {Jr.},
  {Pier}, {Pope}, {Pourbaix}, {Quinn}, {Raddick}, {Fiorentin}, {Richards},
  {Richmond}, {Rix}, {Rockosi}, {Schlegel}, {Schneider}, {Scranton}, {Selja},
  {Sheldon}, {Shimasaku}, {Silvestri}, {Smith}, {Smolcic}, {Snedden},
  {Stoughton}, {Strauss}, {SubbaRao1}, {Suto}, {Szalay}, {Szapudi}, {Szkody},
  {Tegmark}, {Thakar}, {Tremonti}, {Tucker}, {Uomoto}, {Berk}, {Vandenberg},
  {Vidrih}, {Vogeley}, {Voges}, {Vogt}, {Weinberg}, {West}, {White}, {Wilhite},
  {Yanny}, {Yocum}, {York}, {Zehavi}, {Zibetti}, \& {Zucker}}]{sdssdr5}
{Adelman-McCarthy}, J.~K., {Agueros}, M.~A., {Allam}, S.~S., {et~al.} 2007, in
  press

\bibitem[{{Akerlof} {et~al.}(2003){Akerlof}, {Kehoe}, {McKay}, {Rykoff},
  {Smith}, {Casperson}, {McGowan}, {Vestrand}, {Wozniak}, {Wren}, {Ashley},
  {Phillips}, {Marshall}, {Epps}, \& {Schier}}]{akerlof03}
{Akerlof}, C.~W., {Kehoe}, R.~L., {McKay}, T.~A., {et~al.} 2003, \pasp, 115,
  132

\bibitem[{{Anderson} {et~al.}(2007){Anderson}, {Margon}, {Voges}, {Plotkin},
  {Syphers}, {Haggard}, {Collinge}, {Meyer}, {Strauss}, {Ag{\"u}eros}, {Hall},
  {Homer}, {Ivezi{\'c}}, {Richards}, {Richmond}, {Schneider}, {Stinson},
  {Vanden Berk}, \& {York}}]{anderson2007}
{Anderson}, S.~F., {Margon}, B., {Voges}, W., {et~al.} 2007, \aj, 133, 313

\bibitem[{{Barbon} {et~al.}(1982){Barbon}, {Ciatti}, \& {Rosino}}]{barbon1982}
{Barbon}, R., {Ciatti}, F., \& {Rosino}, L. 1982, \aap, 116, 35

\bibitem[{{Bloom} {et~al.}(1999){Bloom}, {Kulkarni}, {Djorgovski},
  {Eichelberger}, {Cote}, {Blakeslee}, {Odewahn}, {Harrison}, {Frail},
  {Filippenko}, {Leonard}, {Riess}, {Spinrad}, {Stern}, {Bunker}, {Dey},
  {Grossan}, {Perlmutter}, {Knop}, {Hook}, \& {Feroci}}]{bloom99}
{Bloom}, J.~S., {Kulkarni}, S.~R., {Djorgovski}, S.~G., {et~al.} 1999, \nat,
  401, 453

\bibitem[{{Branch} {et~al.}(1981){Branch}, {Falk}, {Uomoto}, {Wills}, {McCall},
  \& {Rybski}}]{branch1981}
{Branch}, D., {Falk}, S.~W., {Uomoto}, A.~K., {et~al.} 1981, \apj, 244, 780

\bibitem[{{Buta}(1982)}]{buta1982}
{Buta}, R.~J. 1982, \pasp, 94, 578

\bibitem[{{Condon} {et~al.}(1998){Condon}, {Cotton}, {Greisen}, {Yin},
  {Perley}, {Taylor}, \& {Broderick}}]{condon1998}
{Condon}, J.~J., {Cotton}, W.~D., {Greisen}, E.~W., {et~al.} 1998, \aj, 115,
  1693

\bibitem[{{Fassia} {et~al.}(2001){Fassia}, {Meikle}, {Chugai}, {Geballe},
  {Lundqvist}, {Walton}, {Pollacco}, {Veilleux}, {Wright}, {Pettini}, {Kerr},
  {Puchnarewicz}, {Puxley}, {Irwin}, {Packham}, {Smartt}, \&
  {Harmer}}]{fassia2001}
{Fassia}, A., {Meikle}, W.~P.~S., {Chugai}, N., {et~al.} 2001, \mnras, 325, 907

\bibitem[{{Filippenko}(1997)}]{filippenko1997}
{Filippenko}, A.~V. 1997, \araa, 35, 309

\bibitem[{{Fruchter} {et~al.}(2006){Fruchter}, {Levan}, {Strolger},
  {Vreeswijk}, {Thorsett}, {Bersier}, {Burud}, {Castro Cer{\'o}n},
  {Castro-Tirado}, {Conselice}, {Dahlen}, {Ferguson}, {Fynbo}, {Garnavich},
  {Gibbons}, {Gorosabel}, {Gull}, {Hjorth}, {Holland}, {Kouveliotou}, {Levay},
  {Livio}, {Metzger}, {Nugent}, {Petro}, {Pian}, {Rhoads}, {Riess}, {Sahu},
  {Smette}, {Tanvir}, {Wijers}, \& {Woosley}}]{fruchter2006}
{Fruchter}, A.~S., {Levan}, A.~J., {Strolger}, L., {et~al.} 2006, \nat, 441,
  463

\bibitem[{{Gaskell}(1992)}]{gaskell1992}
{Gaskell}, C.~M. 1992, \apjl, 389, L17

\bibitem[{{Hill} {et~al.}(1998){Hill}, {Nicklas}, {MacQueen}, {Tejada}, {Cobos
  Duenas}, \& {Mitsch}}]{hill1998}
{Hill}, G.~J., {Nicklas}, H.~E., {MacQueen}, P.~J., {et~al.} 1998, in Proc.
  SPIE Vol. 3355, p. 375-386, Optical Astronomical Instrumentation, Sandro
  D'Odorico; Ed., 375--386

\bibitem[{{H{\"o}flich} {et~al.}(1999){H{\"o}flich}, {Wheeler}, \&
  {Wang}}]{hww1999}
{H{\"o}flich}, P., {Wheeler}, J.~C., \& {Wang}, L. 1999, \apj, 521, 179

\bibitem[{{Jeffery} \& {Branch}(1990)}]{jeffery_branch1990}
{Jeffery}, D.~J. \& {Branch}, D. 1990, in Supernovae, Jerusalem Winter School
  for Theoretical Physics, ed. J.~C. {Wheeler}, T.~{Piran}, \& S.~{Weinberg},
  149--+

\bibitem[{{Lazzati} {et~al.}(2001){Lazzati}, {Covino}, {Ghisellini}, {Fugazza},
  {Campana}, {Saracco}, {Price}, {Berger}, {Kulkarni}, {Ramirez-Ruiz},
  {Cimatti}, {Della Valle}, {di Serego Alighieri}, {Celotti}, {Haardt},
  {Israel}, \& {Stella}}]{lazzati01}
{Lazzati}, D., {Covino}, S., {Ghisellini}, G., {et~al.} 2001, \aap, 378, 996

\bibitem[{{MacFadyen} {et~al.}(2001){MacFadyen}, {Woosley}, \&
  {Heger}}]{macfadyen2001}
{MacFadyen}, A.~I., {Woosley}, S.~E., \& {Heger}, A. 2001, \apj, 550, 410

\bibitem[{{Maeda} {et~al.}(2007){Maeda}, {Tanaka}, {Nomoto}, {Tominaga},
  {Kawabata}, {Mazzali}, {Umeda}, {Suzuki}, \& {Hattori}}]{maeda2007}
{Maeda}, K., {Tanaka}, M., {Nomoto}, K., {et~al.} 2007, ArXiv e-prints, 705

\bibitem[{{Nakar} \& {Piran}(2003)}]{nakar_piran2003}
{Nakar}, E. \& {Piran}, T. 2003, New Astronomy, 8, 141

\bibitem[{{Ofek} {et~al.}(2007){Ofek}, {Cameron}, {Kasliwal}, {Gal-Yam}, {Rau},
  {Kulkarni}, {Frail}, {Chandra}, {Cenko}, {Soderberg}, \& {Immler}}]{ofek2007}
{Ofek}, E.~O., {Cameron}, P.~B., {Kasliwal}, M.~M., {et~al.} 2007, \apjl, 659,
  L13

\bibitem[{{Oke} {et~al.}(1995){Oke}, {Cohen}, {Carr}, {Cromer}, {Dingizian},
  {Harris}, {Labrecque}, {Lucinio}, {Schaal}, {Epps}, \& {Miller}}]{oke1995}
{Oke}, J.~B., {Cohen}, J.~G., {Carr}, M., {et~al.} 1995, \pasp, 107, 375

\bibitem[{{Perlmutter} {et~al.}(1999){Perlmutter}, {Aldering}, {Goldhaber},
  {Knop}, {Nugent}, {Castro}, {Deustua}, {Fabbro}, {Goobar}, {Groom}, {Hook},
  {Kim}, {Kim}, {Lee}, {Nunes}, {Pain}, {Pennypacker}, {Quimby}, {Lidman},
  {Ellis}, {Irwin}, {McMahon}, {Ruiz-Lapuente}, {Walton}, {Schaefer}, {Boyle},
  {Filippenko}, {Matheson}, {Fruchter}, {Panagia}, {Newberg}, {Couch}, \& {The
  Supernova Cosmology Project}}]{perlmutter1999}
{Perlmutter}, S., {Aldering}, G., {Goldhaber}, G., {et~al.} 1999, \apj, 517,
  565

\bibitem[{{Quimby} {et~al.}(2007){Quimby}, {Castro}, \&
  {Mondol}}]{SN2006tf_iauc}
{Quimby}, R., {Castro}, F., \& {Mondol}, P. 2007, \iaucirc, 8790, 2

\bibitem[{{Quimby}(2006)}]{quimby_phd}
{Quimby}, R.~M. 2006, PhD thesis, University of Texas, United States -- Texas

\bibitem[{{Schlegel} {et~al.}(1998){Schlegel}, {Finkbeiner}, \&
  {Davis}}]{schlegel1998}
{Schlegel}, D.~J., {Finkbeiner}, D.~P., \& {Davis}, M. 1998, \apj, 500, 525

\bibitem[{{Smith} {et~al.}(2006){Smith}, {Li}, {Foley}, {Wheeler}, {Pooley},
  {Chornock}, {Filippenko}, {Silverman}, {Quimby}, {Bloom}, \&
  {Hansen}}]{smith2007}
{Smith}, N., {Li}, W., {Foley}, R.~J., {et~al.} 2006, ArXiv Astrophysics
  e-prints

\bibitem[{{Stanek} {et~al.}(2003){Stanek}, {Matheson}, {Garnavich}, {Martini},
  {Berlind}, {Caldwell}, {Challis}, {Brown}, {Schild}, {Krisciunas}, {Calkins},
  {Lee}, {Hathi}, {Jansen}, {Windhorst}, {Echevarria}, {Eisenstein}, {Pindor},
  {Olszewski}, {Harding}, {Holland}, \& {Bersier}}]{stanek03}
{Stanek}, K.~Z., {Matheson}, T., {Garnavich}, P.~M., {et~al.} 2003, \apjl, 591,
  L17

\bibitem[{{Voges} {et~al.}(1999){Voges}, {Aschenbach}, {Boller},
  {Br{\"a}uninger}, {Briel}, {Burkert}, {Dennerl}, {Englhauser}, {Gruber},
  {Haberl}, {Hartner}, {Hasinger}, {K{\"u}rster}, {Pfeffermann}, {Pietsch},
  {Predehl}, {Rosso}, {Schmitt}, {Tr{\"u}mper}, \& {Zimmermann}}]{voges1999}
{Voges}, W., {Aschenbach}, B., {Boller}, T., {et~al.} 1999, \aap, 349, 389

\bibitem[{{Voges} {et~al.}(2000){Voges}, {Aschenbach}, {Boller}, {Brauninger},
  {Briel}, {Burkert}, {Dennerl}, {Englhauser}, {Gruber}, {Haberl}, {Hartner},
  {Hasinger}, {Pfeffermann}, {Pietsch}, {Predehl}, {Schmitt}, {Trumper}, \&
  {Zimmermann}}]{voges2000}
---. 2000, VizieR Online Data Catalog, 9029, 0

\bibitem[{{White} {et~al.}(1997){White}, {Becker}, {Helfand}, \&
  {Gregg}}]{white1997}
{White}, R.~L., {Becker}, R.~H., {Helfand}, D.~J., \& {Gregg}, M.~D. 1997,
  \apj, 475, 479

\bibitem[{{Wurtz} {et~al.}(1996){Wurtz}, {Stocke}, \& {Yee}}]{wurtz1996}
{Wurtz}, R., {Stocke}, J.~T., \& {Yee}, H.~K.~C. 1996, \apjs, 103, 109

\bibitem[{{Young} {et~al.}(2005){Young}, {Smith}, \& {Johnson}}]{young2005}
{Young}, T.~R., {Smith}, D., \& {Johnson}, T.~A. 2005, \apjl, 625, L87

\end{thebibliography}

\end{document}